# Non-linear Temperature Oscillations in the Plasma Centre on Tore Supra and their Interplay with MHD

V.S. Udintsev, G. Giruzzi, F. Imbeaux, J.-F. Artaud, X. Garbet, G. Huysmans, P. Maget, J.-L. Ségui, A. Bécoulet, G.T. Hoang, E. Joffrin, X. Litaudon, B. Saoutic, and the Tore Supra Team

*Association Euratom-CEA, CEA/DSM/DRFC, CEA/Cadarache,
F-13108 St. Paul-lez-Durance, France*

ABSTRACT. Regular oscillations of the central electron temperature have been observed by means of ECE and SXR diagnostics during non-inductively driven discharges on Tore Supra. These oscillations are sustained by LHCD, do not have a helical structure and, therefore, cannot be ascribed as MHD phenomena. The most probable explanation of this oscillating regime (O-regime) is the assumption that the plasma current density (and, thus, the $q$-profile) and the electron temperature evolve as a non-linearly coupled predator-pray system. The integrated modelling code CRONOS has been used to demonstrate that the coupled heat transport and resistive diffusion equations admit solutions for the electron temperature and the current density which have a cyclic behaviour. Recent experimental results in which the O-regime co-exists with MHD modes will be presented. Because both phenomena are linked to details of the $q$-profile, some interplay between MHD and oscillations may occur. The localisation of magnetic islands allows to obtain an accurate picture of the $q$-profile in the plasma core. In some case, MHD-driven reconnection helps in maintaining a weakly inverted $q$-profile that is found to be, in the CRONOS simulations, a necessary condition to trigger the oscillations.

## 1. Introduction

Studies of nonlinear electron temperature oscillations in the plasma center are conducted on Tore Supra in steady-state plasma discharges. These oscillations do not have any helical structure, their frequency is low, compared to usual Magnetohydrodynamic (MHD) phenomena and, thus, cannot be ascribed to any known MHD instability. The regime in which they are present is called O-regime. First experimental evidence of $T_e$-oscillations, as well as an explanation that involves a nonlinear evolution of the plasma current density and the electron temperature, has been introduced in [1].

Though not MHD by themselves, oscillations may co-exist with MHD modes. In certain situations, this aids in identification of rational $q$-surfaces correctly. In some cases, however, oscillations and MHD are linked more closely to each other, because modes can perturb the $q$-profile in the plasma core. In this paper, recent experimental results concerning the interplay between $T_e$-oscillations and MHD are presented.

## 2. Diagnostics used for analysis

For the analysis of the interplay between nonlinear oscillations and MHD, the following diagnostics on Tore Supra ($R_0 = 2.40$ m, $a = 0.72$ m, $B_T \approx 3.8$ T, circular cross-section tokamak with superconducting magnetic coils) have been used:

1. an Electron Cyclotron Emission (ECE) radiometer [2] that has been recently upgraded to have 1 $GHz$ spaced, 500 $MHz$ bandwidth 32 measuring channels (frequency range 78-110 $GHz$ for the first harmonic O-mode, 94-126.5 $GHz$ for the second harmonic X-mode), operating both in slow (1 $ms$ sampling rate) and fast (10 $\mu s$ sampling rate) acquisition regimes;

2. a set of fast/slow soft X-ray (SXR) cameras (21 horizontal and 37 vertical viewing chords) [3];
3. a hard X-ray (HXR) system [4], as well as fast magnetics (Mirnov coils) and an interferometer/polarimeter diagnostic.

### 3. O-regime. Model for $T_e$-oscillations

Nonlinear electron temperature oscillations have been observed in the following plasma scenarios [1]: plasma current $I_p = 0.2 - 0.7$ *MA*, central density and electron temperature $n_e^0 = 1.5 - 3 \times 10^{19}$ $m^{-3}$ and $T_e^0 = 4 - 6$ *keV*, respectively. The loop voltage was either exactly zero, or very small (< 100 *mV*). The current was generated by means of Lower Hybrid (LH) waves, launched by two couplers with power spectra peaked at $n_{//} = 1.8 - 2.0$ and a total power ≤ 3 *MW*. A typical example of a shot with oscillations is shown in Fig. 1(a). The onset of the O-regime starts during a slow density ramp down at about 13 *s* and exhibits itself as a fast increase of the central electron temperature, typical of the transition to the hot core Lower Hybrid Enhanced Performance (LHEP) mode, often observed during Lower Hybrid Current Drive (LHCD) in Tore Supra [5]. Figures 1(b, c) show a typical evolution of the central ECE and SXR signals and an ECE contour plot for a selected time interval, respectively.

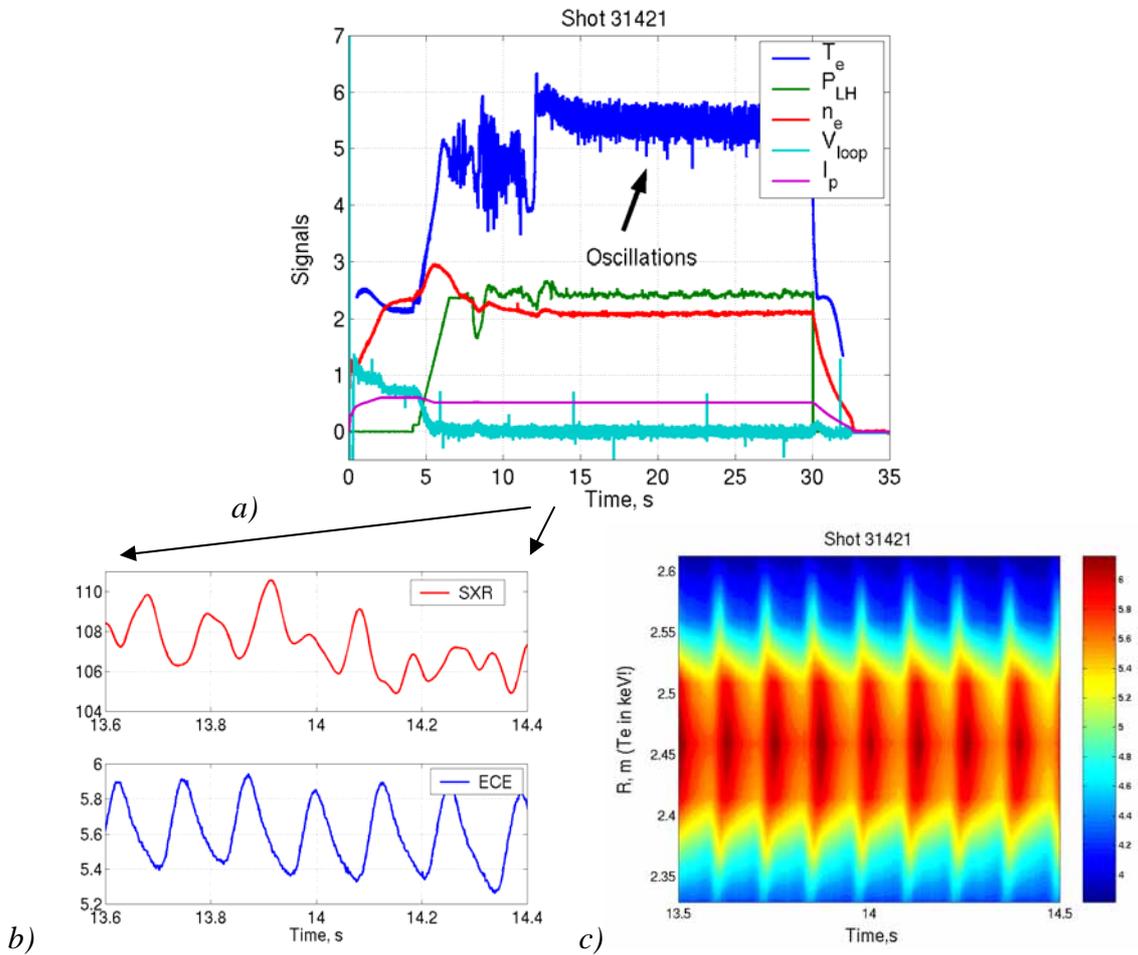

*Figure 1. Time traces (a), evolution of the central ECE and SXR signals (b) and the temperature contour plot (c) for a typical shot with oscillations. Note that there is no phase shift between oscillations at the LFS and the HFS (c), implying no helical structure (typical for MHD phenomena) for these temperature oscillations. Red color in (c) corresponds to higher electron temperature values.*

It has been suggested in [1] that the existence of nonlinear temperature oscillations depends strongly on the evolution of the current density profile. Thus, the oscillation amplitude is found to decrease with the loop voltage and to increase with the radial width of the HXR emission profiles, which have approximately the same shape as the LH driven current profile. The $T_e$-oscillations can be triggered by tailoring the $q$-profile in the plasma core by means of co- and counter-Electron Cyclotron Current Drive (ECCD). In the LHEP mode, the electron transport is much reduced in the plasma core region (typically, at the normalized minor radius $\rho < 0.2 - 0.3$), due to turbulence suppression associated with a negative magnetic shear. All these observations suggest that the origin of the oscillation is linked to the interplay between the current density profile and electron heat transport.

The most plausible explanation of the O-regime is based on the fact that the plasma current density, $j$, and the electron temperature, $T_e$, evolve as a non-linear coupled predator-prey system [6]. At the onset of a transport barrier, the electron heat diffusivity depends on the $q$-profile. The current sources (LHCD, bootstrap, and inductive current), in their own turn, depend on the electron temperature profile. Assuming such dependences, the plasma transport equations can be written as an oversimplified system of 0D equations by reducing the diffusion operator and the source terms to damping/growth terms:

$$\frac{dT_e}{dt} = v_T T_e (1 - \alpha j), \quad (1)$$

$$\frac{dj}{dt} = -v_j j (1 - \beta T_e), \quad (2)$$

This system (Eqs. 1-2) is known as the Lotka-Volterra equations [6-8], which describe the coupled evolution of predator and prey populations living on the same territory, and notoriously admit periodic solutions. In the Eq. 1 describing the evolution of the electron temperature $T_e$, the growth term $v_T$ is positive in order to mimic the transition towards an increased confinement state, while the $-\alpha j$ damping term drives back to a low confinement state when the current profile ceases to be appropriate for stabilising the turbulence. In the Eq. 2, the damping coefficients $-v_j$ corresponds to the current diffusion, while the positive cross-term $\beta T_e$ represents the increase of local current sources with electron temperature. Using the predator-prey system as a guidance, an investigation of models which can produce similar periodic solutions, were conducted in the framework of the full 1D transport equations for current and electron heat in a tokamak plasma. For this purpose, the CRONOS integrated modelling code [9] has been used, with plasma parameters corresponding to a specific O-regime shot. Figure (2) shows the result of CRONOS simulations in which the effect of magnetic shear is involved and $j_{LH}(\rho) \propto T_e(\rho).j(\rho)$. The "shear function" of the Bohm - gyro-Bohm model, which reduces transport when shear s ≤ 0, has been used [10]. The magnetic shear is modulated, and the variations of $T_e$ are due to a propagation of the minimum magnetic shear from $\rho = 0.1$ towards the center.

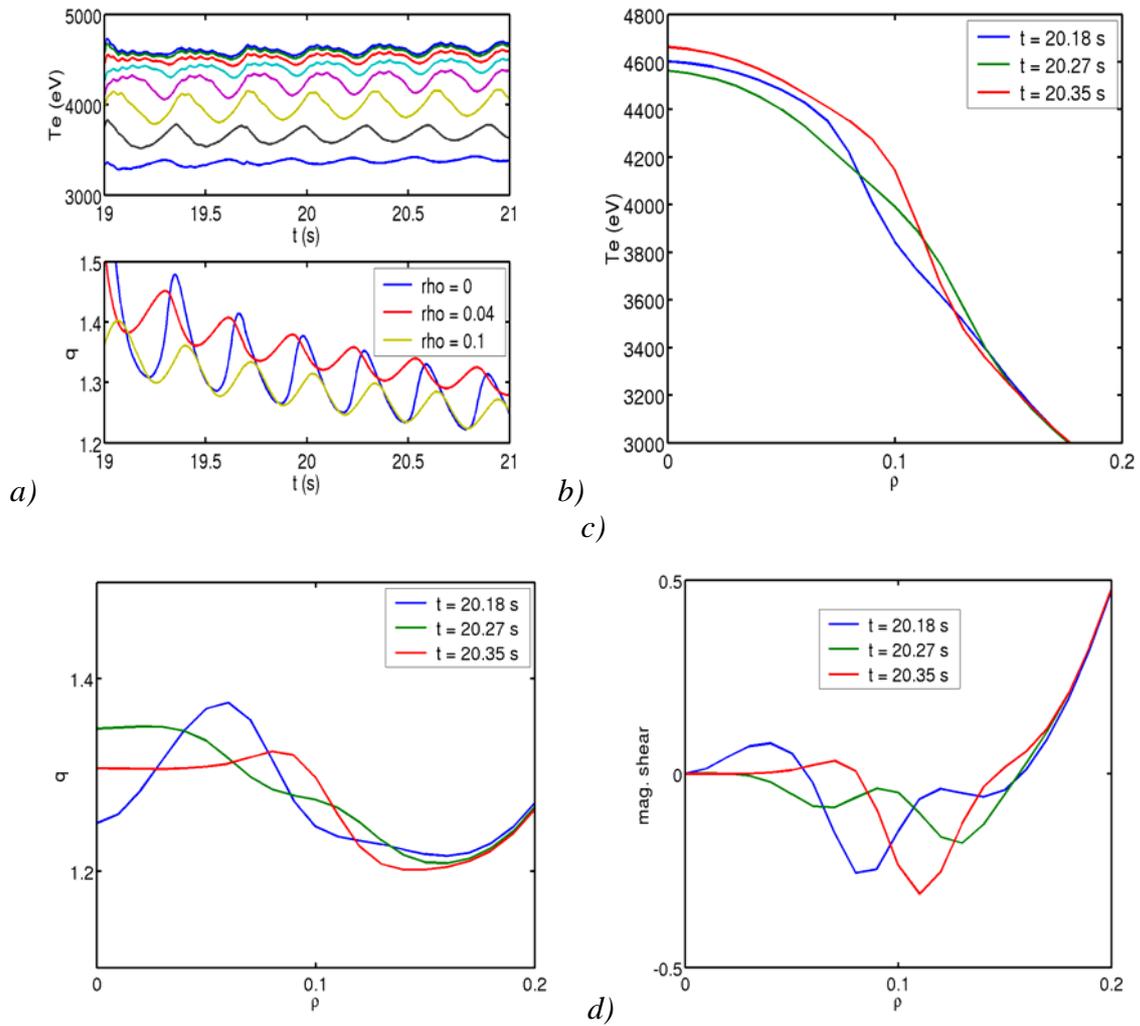

*Figure 2.* CRONOS simulations of nonlinear temperature oscillations. Radial profiles for $T_e$, $q$ and magnetic shear at different times are shown in (b), (c) and (d), respectively.

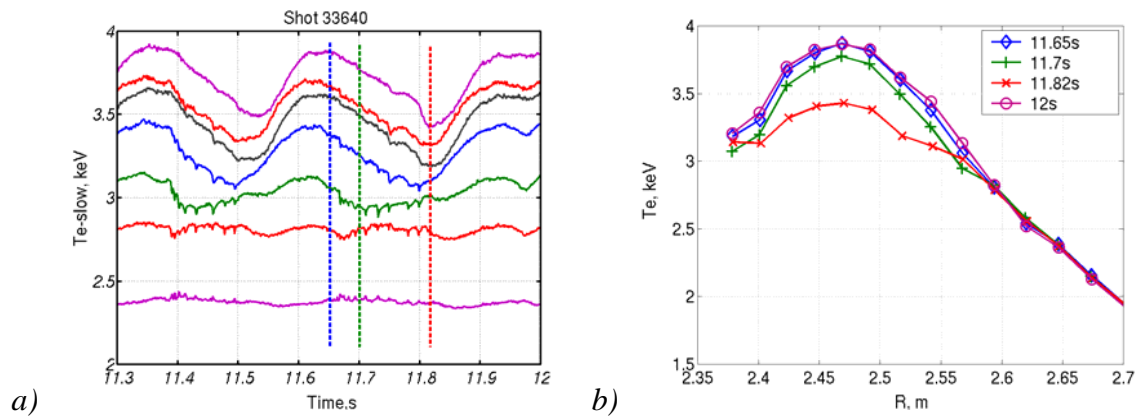

*Figure 3.* Experimental $T_e$-profiles (b) for the shot with oscillations (a). During temperature decrease, a partial flattening of the profile in the ring-shaped plasma region at $r/a = 0.1 – 0.15$ occurs, indicating a change in local plasma confinement.

Temperature oscillations produced by the model are generally consistent with experimental observations. Figure (3) gives experimentally measured $T_e$-profiles for the shot with similar condition that have been used for CRONOS simulations. The last profile (at 12 *s*) is taken after the O-regime phase, when the confinement is improved for the central plasma. Therefore, the O-regime may be considered as a regime of the intermediate confinement, being improved during the rise and top phase of $T_e$-oscillation. More results of CRONOS simulations for the O-regime are reported in [11].

**4. $T_e$-oscillations in presence of MHD activity**

Temperature oscillation can coexist with MHD modes. In some cases, identification of *m* and *n* numbers of MHD modes can be very helpful for a proper determination of the position of rational *q* surfaces. An example is shown in Fig. 4. In this shot with LHCD, central temperature oscillations with frequency of about 6 Hz exist simultaneously with a faster MHD mode of 250 Hz. From Mirnov signals, the MHD mode is identified as 4/1. It cannot be identified properly by ECE, mainly due to the downshifted non-thermal electron emission pollution at the LFS. However, there is also another MHD mode that is strongly coupled to the *m*/*n* = 4/1 one. From ECE, its position at the LFS is determined to be around *R* = 2.6 *m*. Because oscillation of this MHD mode both at the LFS and the HFS are in the same phase, it can be concluded that its *m* number is even. Thus, the most probable candidate is the *m*/*n* = 2/1 tearing mode. Calculations of the *q*-profile performed by EFIT [12] show an inverted *q*-profile, typical of shots with LHCD and favourable for triggering nonlinear temperature oscillations. The calculated position of *q* = 2 and *q* = 4 surfaces is in good agreement with the experimental observations.

Another kind of interplay between temperature oscillations and MHD exhibits itself as a periodic crash and occurs always at the same time during the oscillation cycle (see Fig. 5). The structure of these crashes, with their typical heat pulses propagation observed on the outermost ECE channels, points to the direction of the magnetic reconnection phenomena that occur in low/reversed magnetic shear scenarios. After the crash, a total flattening of the central temperature may occur. CRONOS calculations have confirmed this assumption, therefore, it is expected that these crashes are caused by the double-tearing 2/1 mode, although relatively small in amplitude and/or fast to be resolved by the slow acquisition ECE. These observations are a clear evidence that the current density profile is indeed evolving periodically (as the temperature), and its shape is different during the $T_e$ increase and decrease phases.

Figure 6 shows a very strong interplay between oscillations and fast MHD modes visible by ECE. Interestingly, 2 to 4 central temperature oscillations are bounded by a fast crash, after which the whole dynamics recover. The frequency of this fast MHD mode is about 150 *Hz*, compared to the $T_e$-oscillations frequency of about 10 *Hz*. Very likely, MHD crashes help to maintain a flat *q*-profile in the plasma center, which has been identified as a critical feature for the triggering of oscillations in the modelling involving the magnetic shear effect.

In several shots, a presence of high-frequency (compared to the $T_e$-oscillations frequency), radially localised structures has been observed by ECE. Figure 7 shows ECE time traces soon after the minor disruption at 212 *s* for the shot with reduced LH power (1 *MW*). Small oscillations on top of slower ones can be caused by small-scale MHD modes with higher *m* and *n* numbers, similar to what have been observed on TEXTOR [13].

Unfortunately, no fast magnetic diagnostics were available to identify the *m* and *n* numbers of these modes.

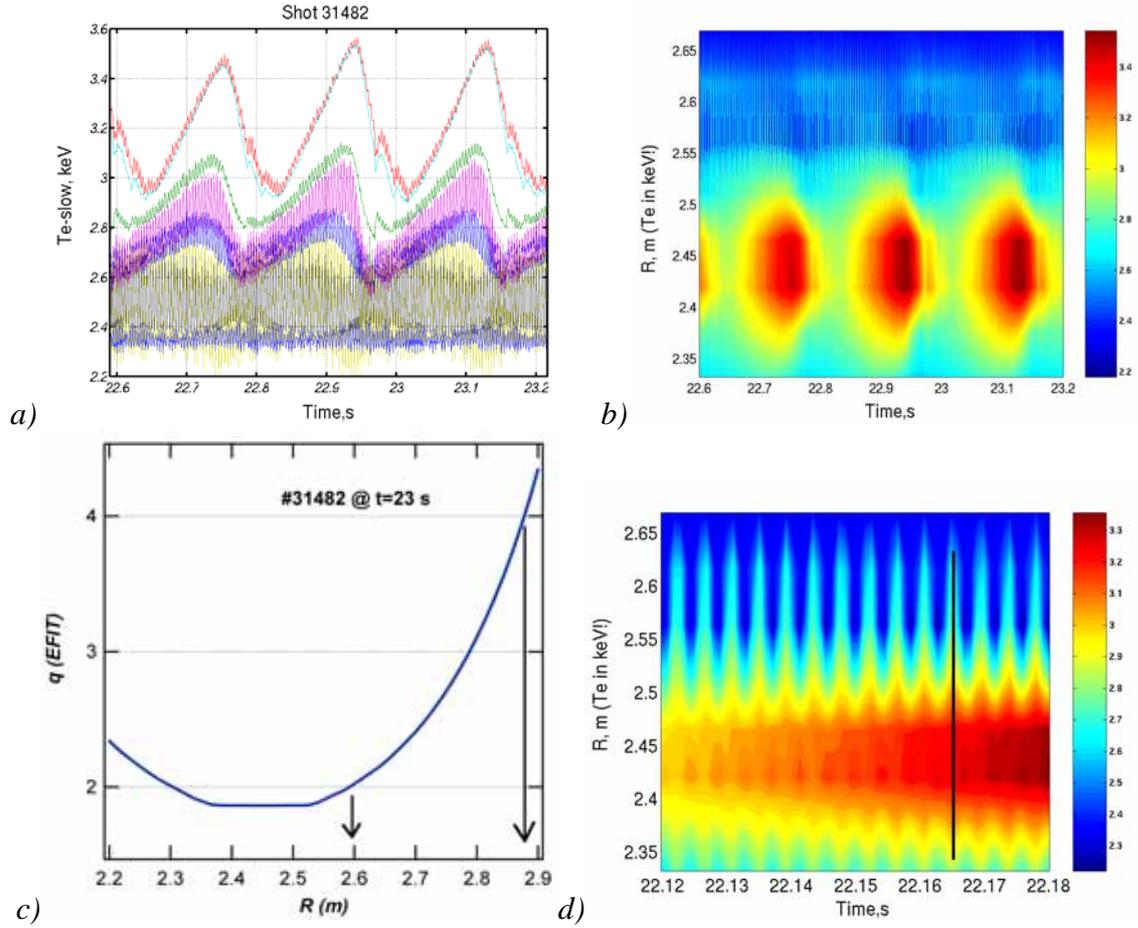

*Figure 4. Temperature oscillations can coexist with MHD modes (a). From the contour plot, the radial position of the mode and, thus, the position of the rational q surface is determined at R = 2.6 m (b, d). Vertical line in (d) shows that there is no change in phase of the mode oscillations between LFS and HFS. This helps to calculate the q-profile for this phase of the shot correctly (c).*

## 5. Conclusions

The regime with central electron temperature oscillations results from a nonlinear coupling between electron heat transport and current diffusion. It likely occurs when turbulence stabilisation near the plasma core starts to develop, but not reaching a complete and stable internal transport barrier (ITB) state. In present Tore Supra experiments, the onset of the $T_e$-oscillations is likely linked to the almost full LH current drive, which introduces a strong dependence of the current diffusion on the temperature and *q* profiles. Nevertheless, this kind of dependence is not unique to LHCD: for example, the bootstrap current also depends on the pressure and *q* profiles in a similar way. Therefore, it might be possible for similar oscillations to exist in very different regimes, for instance in steady-state scenarios with high bootstrap fraction.

Although these temperature oscillations are not of MHD origin, an interplay with MHD modes occurs in several cases, because both phenomena are linked to details in *q*-profile. The localisation of magnetic islands aids to get an accurate picture of the *q*-profile in the plasma core. In some cases, MHD-driven reconnection helps in maintaining a weakly inverted *q*-profile that is, in the CRONOS simulations, a necessary condition to trigger the oscillations.

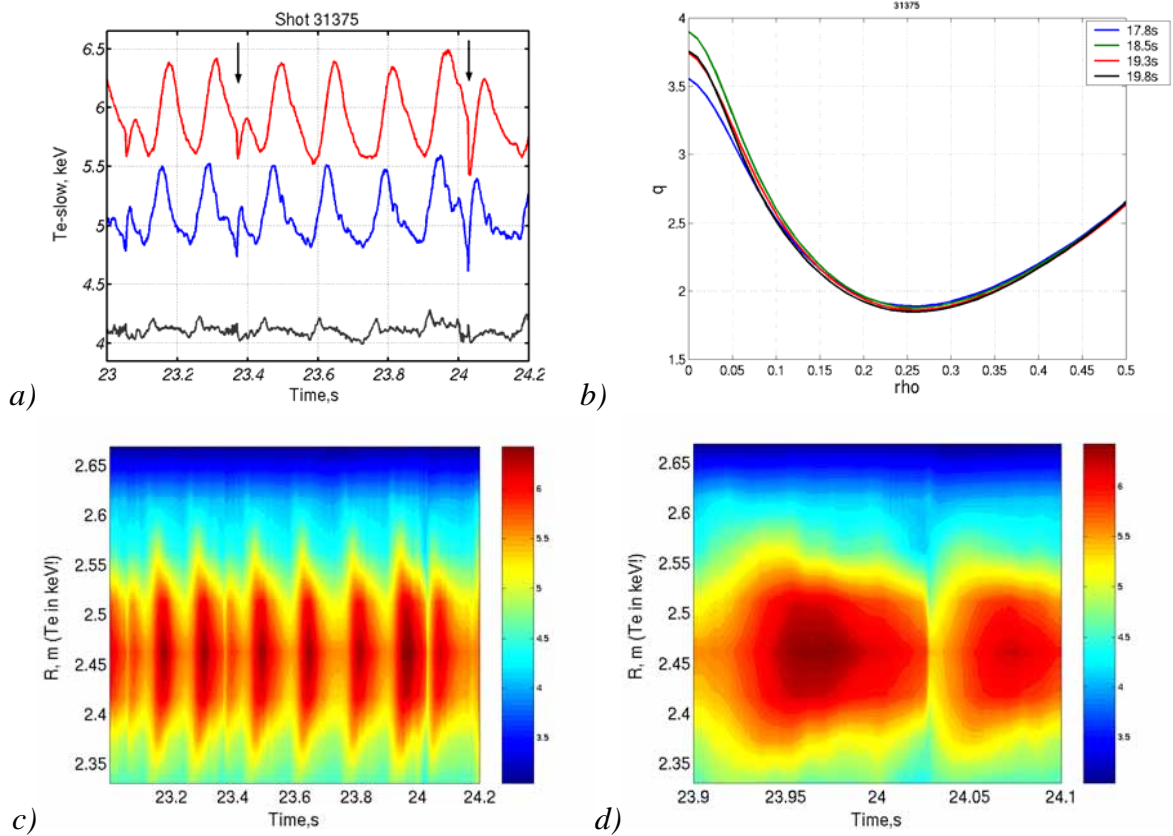

**Figure 5**. At $V_{loop}$ =0, $P_{LH}$ =2.5 MW, a crash-like event (probably due to m/n = 2/1 double tearing mode) appears during O-regime, as indicated by black arrows in (a). After the crash, the temperature flattens in the plasma center (c, d). CRONOS calculations reveal an inverted q-profile in this shot (b).

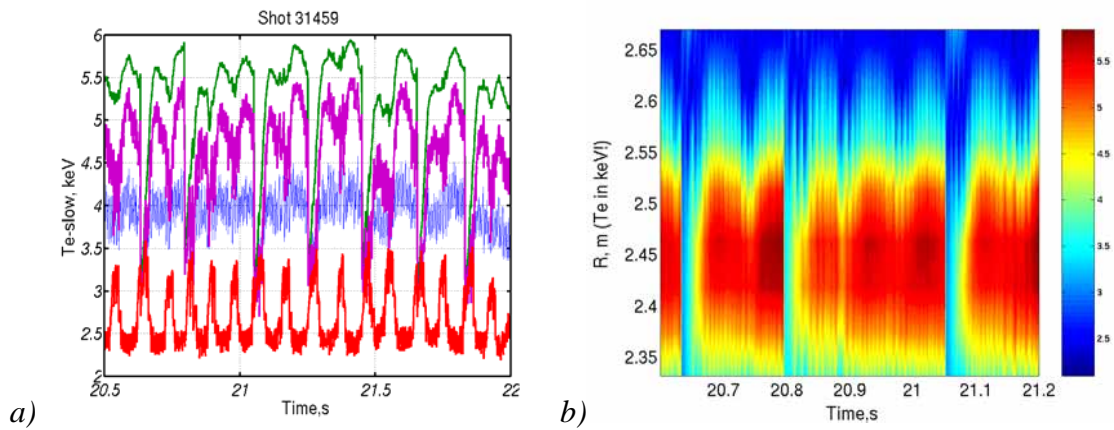

**Figure 6**. The O-regime with 10 Hz oscillations coexist with fast MHD mode (150 Hz) and periodic crashes every 2 to 4 oscillation cycles.

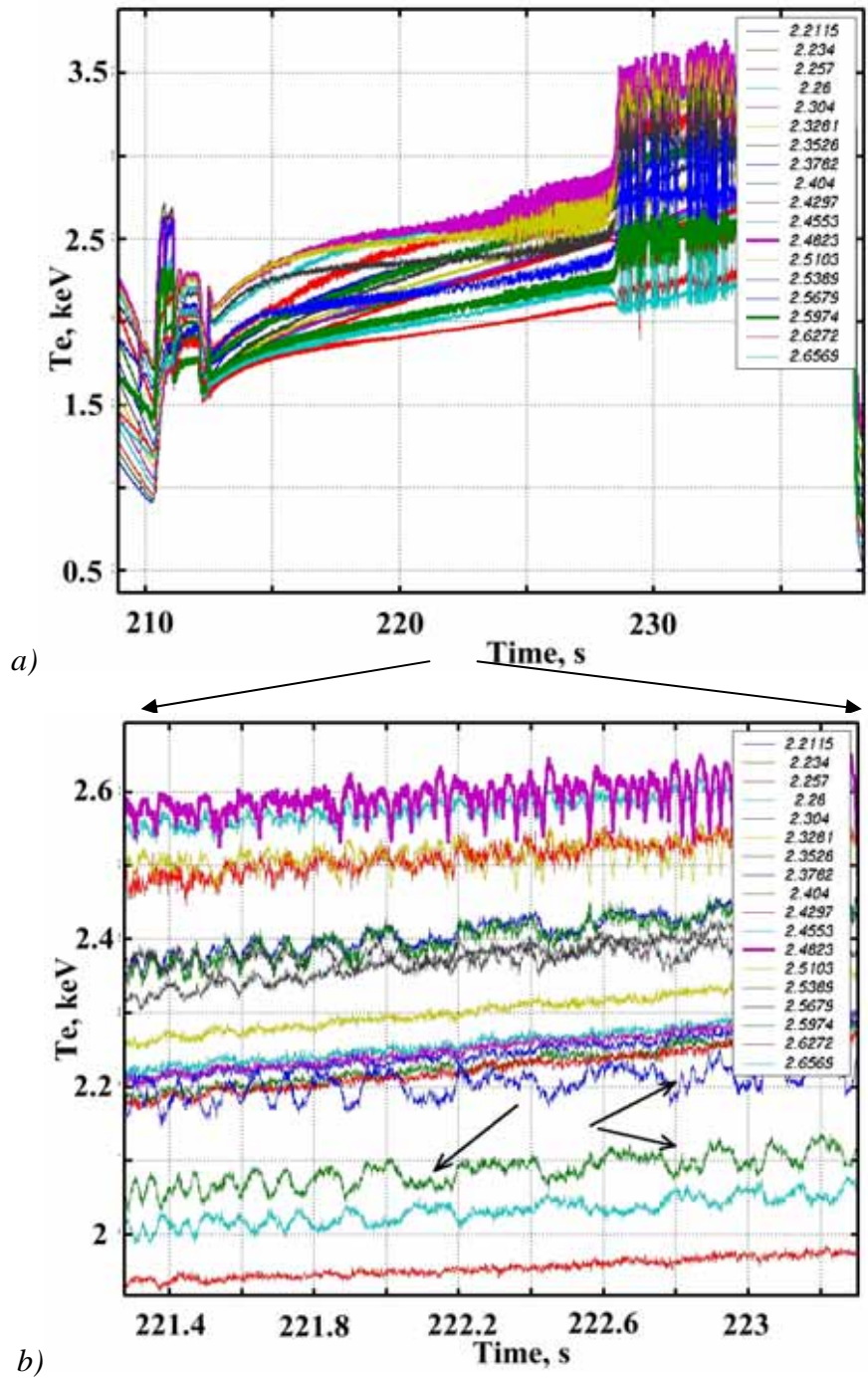

*Figure 7*. *Evidence for the fast MHD modes with (possibly) high m and n numbers (black arrows in (b)) during the O-regime that starts soon after the minor disruption at 212 s (a). For this phase of the shot, LH power has been lowered down to 1 MW at 210 s. The outermost ECE channels (R > 2.62 m) are influenced by the downshifted non-thermal ECE emission.*